\documentclass[prd,aps,preprint,tightenlines,nofootinbib,superscriptaddress]{revtex4}
\usepackage{mathrsfs}
\usepackage{amsfonts}
\usepackage{amsmath}
\usepackage{array}
\usepackage{verbatim}
\usepackage{epsfig}
\usepackage{graphicx}

\begin{document}
\title{Energy Evolution for the Sivers Asymmetries in Hard Processes}

\author{Peng Sun}
\affiliation{Nuclear Science Division, Lawrence Berkeley National
Laboratory, Berkeley, CA 94720, USA}
\author{Feng Yuan}
\affiliation{Nuclear Science Division, Lawrence Berkeley National
Laboratory, Berkeley, CA 94720, USA}


\begin{abstract}
We investigate the energy evolution of the azimuthal spin asymmetries in semi-inclusive
hadron production in deep inelastic scattering (SIDIS) and Drell-Yan lepton pair production
in $pp$ collisions. The scale dependence is evaluated by applying an approximate solution to
the Collins-Soper-Sterman (CSS) evolution equation at one-loop order which is
adequate for moderate $Q^2$ variations. This describes well the unpolarized
cross sections for SIDIS and Drell-Yan process in the $Q^2$ range of 2.4-100GeV$^2$.
A combined analysis of the Sivers asymmetries in SIDIS from HERMES and COMPASS
experiments, and the predictions for the Drell-Yan process at RHIC at $\sqrt{S}=200$GeV are presented.
We further extend to the Collins asymmetries and find, for the first time, a consistent description for
HERMES/COMPASS and BELLE experiments with the evolution effects. We emphasize an important
test of the evolution effects by studying di-hadron azimuthal asymmetry in $e^+e^-$ annihilation
at moderate energy range, such as at BEPC at $\sqrt{S}=4.6$GeV.
\end{abstract}

 \maketitle


{\it Introduction.} Transverse spin azimuthal angular asymmetries in hadronic processes
have attracted great attentions in recent years. This is not only because
the associated observables are keen to provide important information on
nontrivial hadronic structures, but also because they are sensitive to the
strong interaction dynamics~\cite{Boer:2011fh}. The latter involves core feature of quantum
chromodynamics (QCD): the factorization and universality of the associated
parton distributions and fragmentation functions, and the energy evolution
in hard scattering processes.

Among of these observables the major focuses are the Sivers and Collins asymmetries in semi-inclusive
hadron production in deep inelastic scattering (SIDIS) and Drell-Yan lepton
pair production in $pp$ collisions, and di-hadron production in $e^+e^-$
annihilation processes.
The Sivers effects come from the asymmetric transverse momentum
dependent (TMD) parton distribution in nucleon which correlates with
the transverse polarization vector $S_\perp$, whereas the Collins effects
come from the similar correlation in the fragmentation process associated
with the quark polarization. However, these two functions
have different universality properties: Sivers function differs by
sign between SIDIS and Drell-Yan processes~\cite{Brodsky:2002cx,Collins:2002kn},
while the Collins function is universal between SIDIS and $e^+e^-$ processes.
Both asymmetries have been observed in SIDIS from HERMES, COMPASS,
and JLab Hall A experiments~\cite{Adolph:2012sp,Airapetian:2009ae,Airapetian:2010ds,Alekseev:2008aa,Adolph:2012sn}. In addition,
Collins asymmetry has been observed in $e^+e^-$ process by BELLE collaboration~\cite{Seidl:2008xc}.

The experimental test of the above universality, in
particular, for the Sivers asymmetries between Drell-Yan and SIDIS, is
one of top questions in hadronic physics. However, the Sivers asymmetries were
observed in SIDIS with $Q^2$ around $3$GeV$^2$, whereas the Drell-Yan processes
will be measured in the range that is greater than $20$GeV$^2$.
In order to consolidate the universality test, the $Q^2$ dependence
of the Sivers asymmetry must be understood correctly.
The theoretical framework to study the energy evolution of these
observables has been well developed, where the Collins-Soper-Sterman
(CSS) equation~\cite{Collins:1981uk,Collins:1984kg} for both spin-average and single-spin dependent
cross section has been derived~\cite{Boer:2001he,Ji:2004wu,Idilbi:2004vb,Collins,Aybat:2011zv,Aybat:2011ge}.
The CSS formalism has been applied
successfully to describe low transverse momentum distribution of vector
boson (Drell-Yan, W/Z) production in unpolarized $pp$ collisions (see, for example,
Ref.~\cite{Landry:2002ix}). Early estimate for $Q^2$ dependence of
the SSA~\cite{Boer:2001he} was limited to high $Q^2$ range.
A recent calculation found a surprising strong evolution effects from
HERMES/COMPASS energies to typical Drell-Yan energy~\cite{Aybat:2011ta}. This evolution
formalism was later applied in a fit to HERMES/COMPASS data~\cite{Anselmino:2012aa}.
The result of Ref.~\cite{Aybat:2011ta} has raised great concerns in the experimental proposals, since the
predicted asymmetries for Drell-Yan processes would be
too small due to the evolution.
In this paper, we will examine these studies, and carefully investigate
the $Q^2$ evolution of both spin average and single-spin dependent
cross sections. By doing so, we find that the previous study of Ref.~\cite{Aybat:2011ta} over-estimated
the evolution effects. In particular, the transverse momentum spectrum of
the Drell-Yan process in the relevant $Q^2$ range can not
be described by the TMD quark distributions proposed in
Ref.~\cite{Aybat:2011zv,Aybat:2011ta} (see Fig.~1 below).

In our calculation,  we take an alternative approach,
following the original suggestion of Ref.~\cite{Ji:2004wu,Idilbi:2004vb}, by directly applying the CSS equation
at one-loop order from low to high energies. The one-loop evolution kernel
contains a term which predicts a $P_T$ broadening
effects at higher $Q^2$. We will show that this can describe the transverse momentum
distribution for both SIDIS and Drell-Yan processes, which cover $Q^2$ in the
range of $2.4$-$100$GeV$^2$. We extend the evolution to the Sivers asymmetries
in these processes, and perform a combined fit to the HERMES and COMPASS
data. The predictions
for the SSA in Drell-Yan process at RHIC will be updated with the evolution
effects. Finally, we will apply the evolution equation to the Collins asymmetries
in SIDIS and di-hadron production in $e^+e^-$ annihilation.

{\it Collins-Soper-Sterman Evolution.}
We take the SIDIS as an example, where $e (\ell)+p(P)\to e(\ell') + h (P_h) + X$,
which proceeds through exchange of a virtual photon
with momentum $q_\mu=\ell_\mu-\ell'_\mu$ and invariant mass
$Q^2=-q^2$.
When $P_{h\perp}\ll Q$, the transverse-momentum-dependent
factorization formalism applies, according which the
differential  SIDIS cross section can be written as
\begin{eqnarray}
    \frac{d\sigma(S_\perp)}{dx_Bdydz_hd^2\vec {P}_{h\perp}}
      &=& \sigma_0\times\left[F_{UU}+\epsilon^{\alpha\beta}S_\perp^\alpha
 F_{\rm sivers}^{\beta}
      \right] \ ,
\end{eqnarray}
where $\sigma_0=4\pi\alpha^2_{\rm em}S_{ep}/{Q^4}\times (1-y+y^2/2)
x_B$, and $y$, $x_B$, and $z_h$ are usual kinematics for SIDIS. We
only keep the terms we are interested in: $F_{UU}$ corresponds to
the unpolarized cross section, and $F_{\rm sivers}$ to the Sivers
function contribution to the single-transverse-spin asymmetry.
$F_{UU}$ and $F_{\rm sivers}$ depend on the kinematical variables,
$x_B$, $z_h$, $Q^2$, and $P_{h\perp}$, can be written into a factorization
form with TMD quark distribution and fragmentation functions and soft
and hard factors.
The $Q^2$ dependence of $F_{UU,\rm sivers}$ can be calculated
from perturbative QCD, and is controlled by the CSS
evolution equation, which is easily formulated in the impact parameter
space. For example, for
$F_{\rm sivers}^\alpha(Q;P_{h\perp})=\int \frac{d^2b}{(2\pi)^{2}} e^{i \vec{P}_{h\perp}\cdot \vec{b}/z_h} \widetilde
{F}_{\rm sivers}^\alpha(Q;b)$~\footnote{Here we only keep the dominant term at low $P_{h\perp}$ region,
and neglect higher power correction of $P_{h\perp}/Q$.}, we have~\cite{Collins:1984kg},
\begin{equation}
\widetilde {F}_{\rm sivers}^\alpha(Q;b)=\widetilde
{F}_{\rm sivers}^\alpha(Q_0;b)e^{-{\cal S}_{Sud}(Q,Q_0,b)} \  . \label{evo}
\end{equation}
The perturbative calculable evolution effect has been included in the Sudakov
form factor ${\cal S}_{Sud}$. In the complete CSS resummation, $Q_0$ was set at $1/b$, and
the $b_*$ prescription was introduced: $b_*=b/\sqrt{1+b^2/b_{max}^2}$
to deal with the Landau pole singularity. This necessarily
introduces a non-perturbative form factor~\cite{Collins:1984kg}, which can be
determined by comparison to the experimental data~\cite{Landry:2002ix}.

Alternatively, it was argued in Refs.~\cite{Ji:2004wu,Idilbi:2004vb} that we can avoid the Landau pole
singularity by a direct integration from low to high energy scale,
\begin{eqnarray}
{\cal
S}_{Sud}={2C_F}\int_{Q_0}^{Q}\frac{d\bar\mu}{\bar\mu}
\frac{\alpha_s(\bar\mu)}{\pi}\left[\ln \left(\frac{Q^2}{\bar\mu^2}\right)
+\ln\frac{Q_0^2b^2}{c_0^2}-\frac{3}{2}\right]\ , \label{sud}
\end{eqnarray}
where $c_0=2e^{-\gamma_E}$ with the complete one-loop coefficients from a recent
calculation~\cite{Kang:2011mr}, and both $Q$ and $Q_0$ are chosen
in the perturbative region. Because of the residual log term in the integral,
the above Sudakov form factor is not the complete solution to the CSS evolution.
But, it is a reasonable approximation in the moderate $Q$ and $Q_0$ range,
in particular, between the HERMES/COMPASS and typical Drell-Yan energy regions.
For very large $Q$, such as $W/Z$ boson production in $pp$ collisions,
we have to take into account higher order corrections and back to the CSS
formalism. We notice that the same Sudakov form factor applies to both
$\widetilde{F}_{UU}$ and $\widetilde{F}_{\rm sivers}$ since they share the
same evolution kernel and hard factors in the TMD factorization.
It works for the Drell-Yan lepton pair production in $pp$ collisions as well~\footnote{The difference
in hard factors in the TMD factorization does not affect the evolution equation.}.

\begin{figure}[tbp]
\centering
\includegraphics[width=6.1cm]{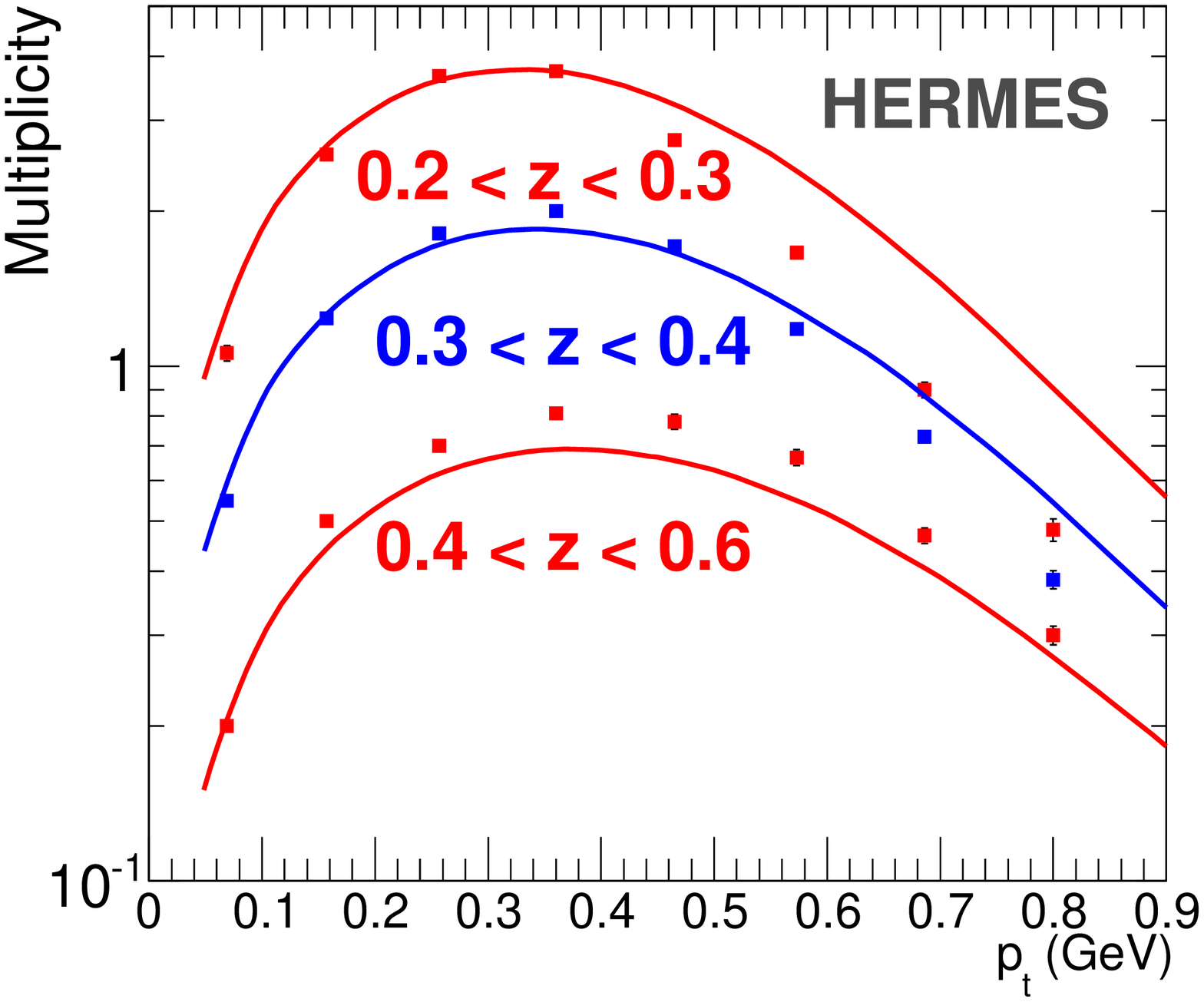}
\includegraphics[width=6.1cm]{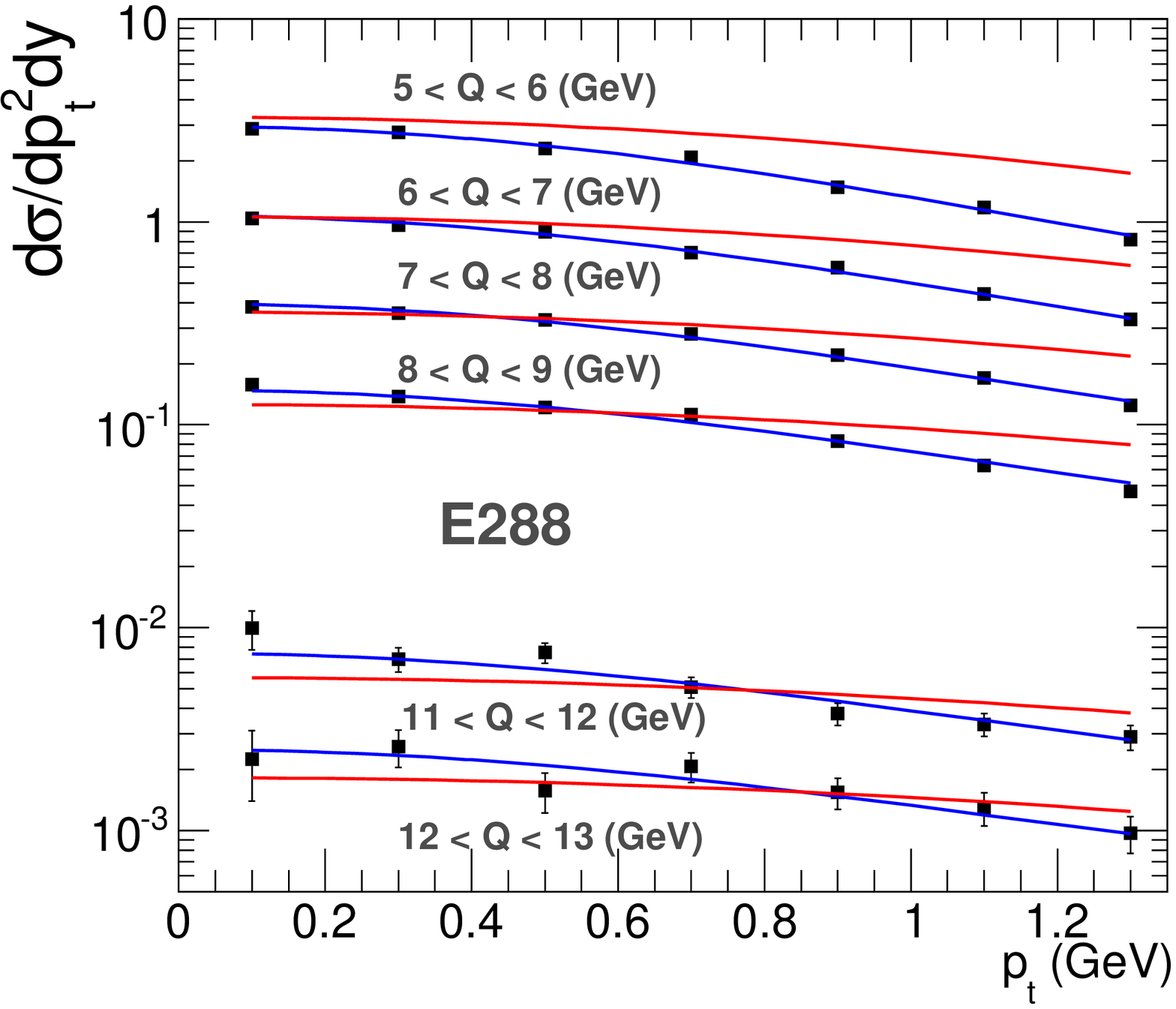}
\caption{Comparison between the theory predictions with
experimental data of the low transverse momentum distribution
of SIDIS at $Q^2=2.4$GeV$^2$ Ref.~\cite{Airapetian:2012ki} and Drell-Yan lepton
pair production in $pp$ collisions with various $Q^2$ range Ref.~\cite{Ito:1980ev}.
The scale evolution comes from the Sudakov form factor of
Eq.~(\ref{sud}). The predictions calculated from the TMD quark distributions of
Ref.~\cite{Aybat:2011zv} with $b_{max}=0.5GeV^{-1}$ and $g_2=0.65GeV^2$
are also shown as red curves in the right panel.}
\label{fig:average}
\end{figure}

Before we study the energy evolution of the SSA, we shall
check the above equation can describe the spin-average cross sections
in the relevant energy range. The majority of the SIDIS data from HERMES and COMPASS
are in a relative low $Q^2$ range. Therefore, we set the lower scale $Q_0$
around these experiments, where a Gaussian assumption
for the TMD quark distribution and fragmentation functions can
well describe the data~\cite{Anselmino:2005ea,Schweitzer:2010tt}.
Translating this into the impact
parameter space, we parameterize ${\widetilde{ F}}_{UU}$ as
\begin{eqnarray}
\widetilde{F}_{UU}(Q_0,b)&=& \sum_q\;e_q^2  \; f_q(x_B)\; D_q(z_h)
e^{-{g_0b^2}-{g_hb^2}/{z_h^2}}   \ ,
\end{eqnarray}
at $Q_0^2=2.4GeV^2$, where $f_q(x_B)$ and $D_q(z_h)$ represent the quark distribution
and fragmentation functions following the CTEQ and DSS set Ref~\cite{deFlorian:2007aj}
at lower scale $Q_0^2=2.4GeV^2$. In the above equation, $g_0$
and $g_h$ represent the transverse momentum dependence coming
from the distribution and fragmentation, respectively. In the left panel of Fig.~1,
we compare the above prediction to the multiplicity distribution in SIDIS
from HERMES experiment~\cite{Airapetian:2012ki} , where we have chosen
$g_0=0.097$ and $g_h=0.045$. These parameters agree well with
those in Ref.~\cite{Schweitzer:2010tt,Aybat:2011ta}.

We can study the $Q^2$ evolution by comparing to the fixed target Drell-Yan process,
with $Q^2$ range from $20$ to 100 GeV$^2$.
To calculate the transverse momentum spectrum for this process,
we apply the universality of the TMD quark distributions, and the evolution
equation from $Q_0$ scale to higher $Q$.
We  plot the comparisons between the theory calculations with the experimental
data in the right panel of Fig.~1. The broadening effects for the Drell-Yan
processes are well reproduced by the evolution effects of Eqs.~(\ref{evo},\ref{sud}).
For comparison, we also plot the predictions from the
TMD quark distributions calculated from Ref.~\cite{Aybat:2011ta} with their
evolution effects.  Clearly, Ref.~\cite{Aybat:2011ta}
over-estimate the broadening effects. It is caused by a modification of
the non-perturbative form factors used in Ref.~\cite{Landry:2002ix} in order to
describe the current SIDIS data, which unfortunately
breaks the original predictions for the Drell-Yan processes in Ref.~\cite{Landry:2002ix}.

{\it Sivers Asymmetries in SIDIS and Drell-Yan.}
Now, we turn to the Sivers single spin asymmetries in SIDIS and Drell-Yan processes.
Similar to the above, we parameterize $\widetilde{F}^\alpha_{\rm sivers}$
at low energy scale $Q_0^2=2.4GeV^2$,
\begin{eqnarray}
\widetilde{F}_{\rm sivers}^\alpha(Q_0,b)&=&  \frac{ib_\perp^\alpha M }{2}
\sum_q\;e_q^2  \; \Delta f_q^{\rm sivers}(x)\;  D_q(z)
e^{-(g_0-g_s)b^2-{g_hb^2}/{z_h^2}}\ ,
\label{siver}
\end{eqnarray}
where $M=0.94GeV$ is a normalization scale, and
we have chosen an additional parameter $g_s$ for transverse momentum dependence
and the fragmentation part remains the same. The function $\Delta f_q(x)= N_q x^{\alpha_q} (1-x)^{\beta_q}
 \frac{(\alpha_q+\beta_q)^{\alpha_q+\beta_q}}{\alpha_q^{\alpha_q} \; \beta_q^{\beta_q}}f_q(x)$
parameterize the $x$-dependence of the quark Sivers function similar to that in Ref.~\cite{Anselmino:2005ea}.
We have the following free parameters: $g_s$, $\alpha_q$, $\beta_q$ and
$N_q$ for valence up, down, and sea quarks.
Since the data are not sufficient to differentiate $g_s$ for
different flavors, we choose the same $g_s$. 
We further assume the same $\beta$ parameter for all quark flavors, and
the same $\alpha$ parameter for all the sea quarks.

\begin{figure}[tbp]
\centering
\includegraphics[width=8.1cm]{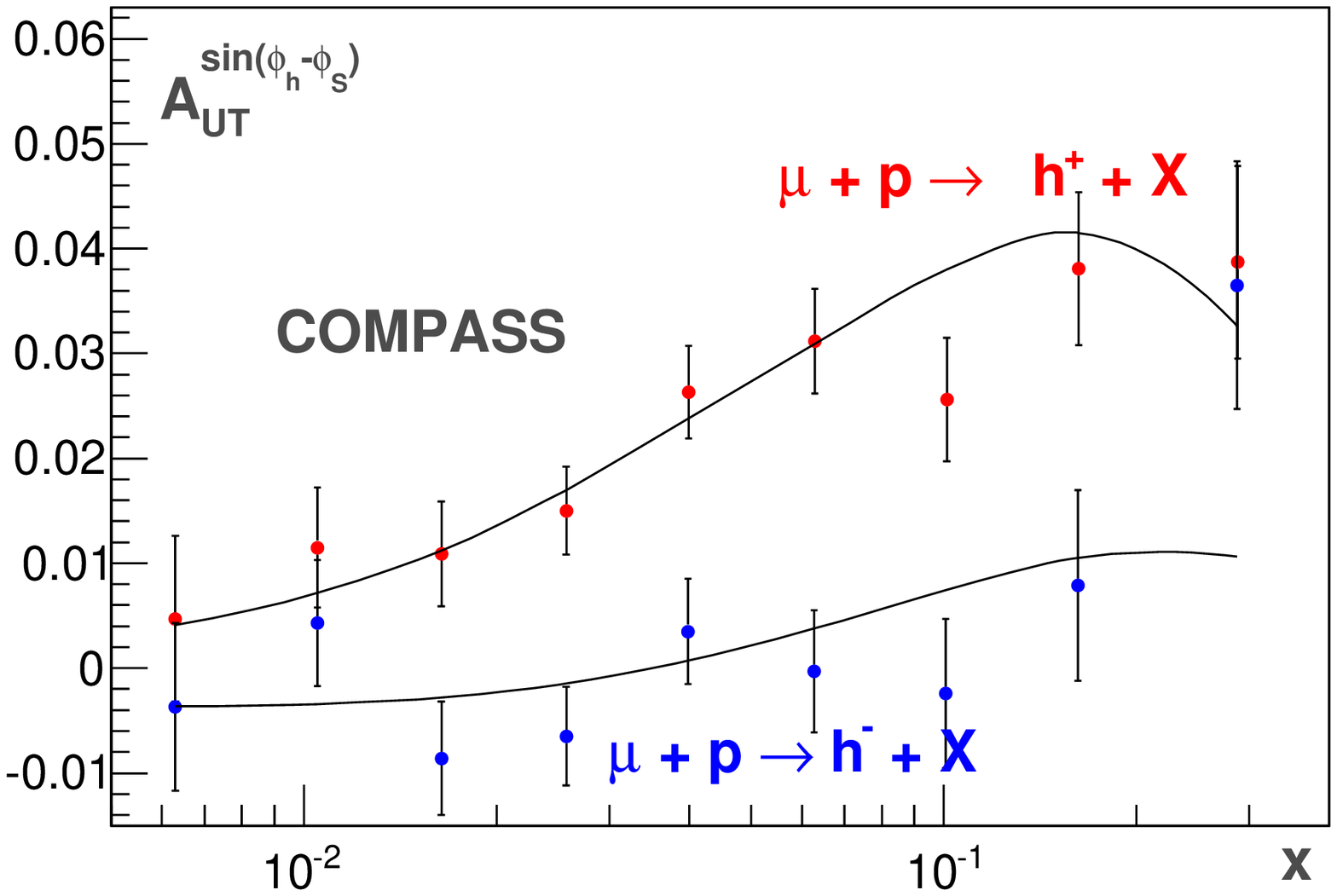}
\includegraphics[width=8.1cm]{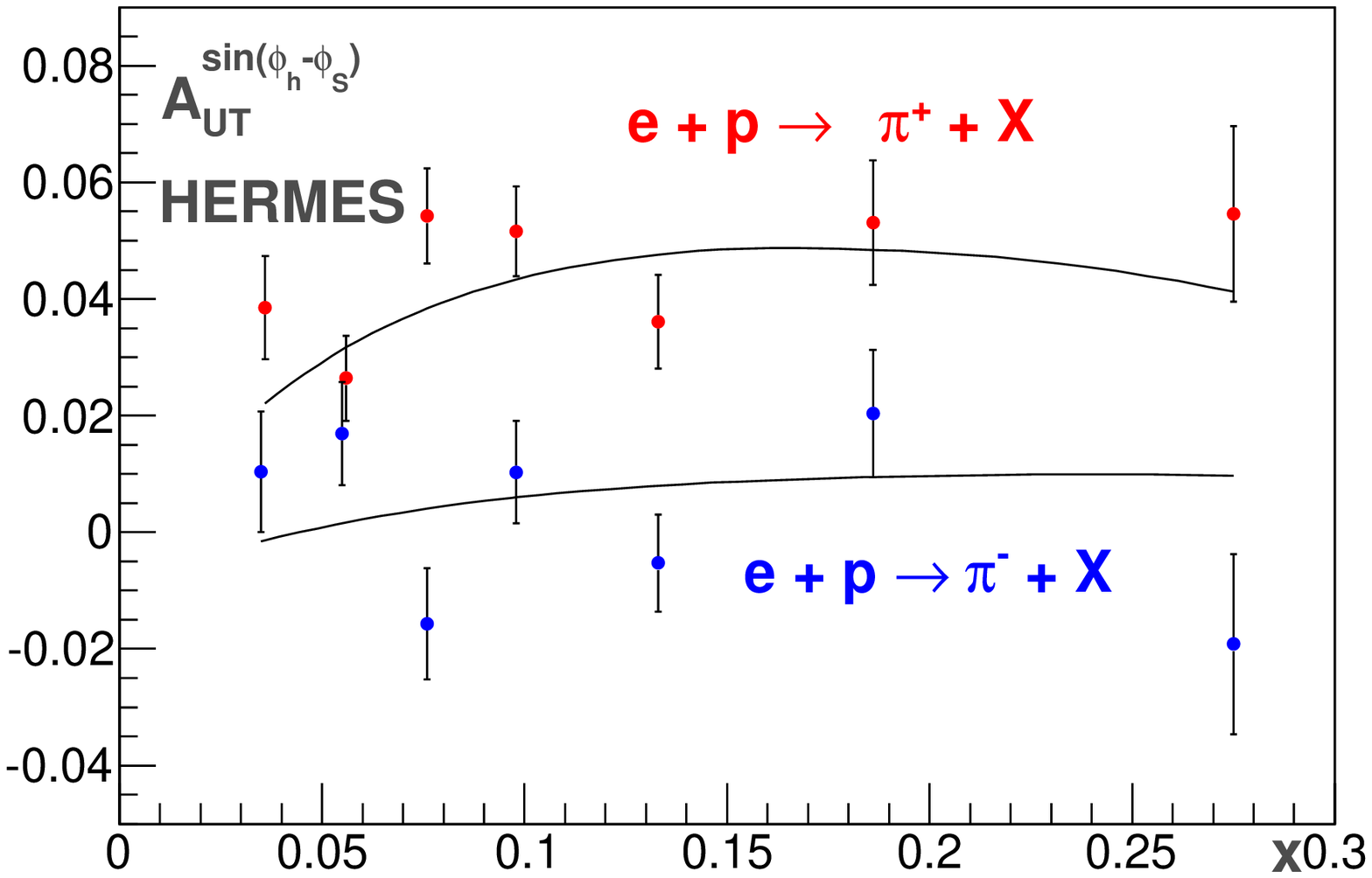}
\caption{Theory fit to the experimental data on Sivers
single spin asymmetries in SIDIS , as functions of $x_B$: left
panel from COMPASS~\cite{Adolph:2012sp} and right from HERMES~\cite{Airapetian:2009ae}. $Q^2$ evolution
has been taken into account from Eq.~(\ref{sud}).}
\label{fig:sivers}
\end{figure}

%
\begin{table}[th!]
\caption{\label{tab:para} Parameters $\{a_i^0\}$
describing our optimum
$\Delta f_i$ in Eq.~(\ref{siver})
at the input scale $Q^2=2.4\,\mathrm{GeV}$.}
\begin{ruledtabular}
\begin{tabular}{ccccc}
flavor $i$ &$N_i$ & $\alpha_i$ & $\beta_i$ & $g_s$ (GeV$^2$) \\
\hline
$u$   & 0.13$\pm$0.023    & 0.81$\pm$0.16  & 4.0$\pm$1.2         & 0.062$\pm$0.005     \\
$d$   & -0.27$\pm$0.12   & 1.41$\pm$0.28  &  4.0$\pm$1.2         & 0.062$\pm$0.005     \\
$s$   & 0.07$\pm$0.06   & 0.58$\pm$0.39  &  4.0$\pm$1.2    & 0.062$\pm$0.005     \\
$\bar{u}$ & -0.07$\pm$0.05    & 0.58$\pm$0.39  & 4.0$\pm$1.2       & 0.062$\pm$0.005     \\
$\bar{d}$ & -0.19$\pm$0.12    & 0.58$\pm$0.39  &  4.0$\pm$1.2      & 0.062$\pm$0.005     \\
\end{tabular}
\end{ruledtabular}
\end{table}

With the above parameterization and the energy evolution effects taken
by Eqs.~(\ref{evo},\ref{sud}) for both spin-average and single-spin-dependent
cross sections, we perform a combined fit to the Sivers asymmetries from HERMES
and COMPASS experiments which scans $Q^2\sim$2.4-$10$GeV$^2$.
We have total of 255 data points, with a minimum $\chi^2$ fit.
The best fit results into $\chi^2/d.o.f=1.08$ and the parameters listed in Table I. As an example,
we show in Fig.~2 the comparisons between the theory calculations
and the experimental data as functions of $x_B$ for COMPASS and
HERMES experiments, which demonstrate a consistent description of
both data.

\begin{figure}[tbp]
\centering
\includegraphics[width=8cm]{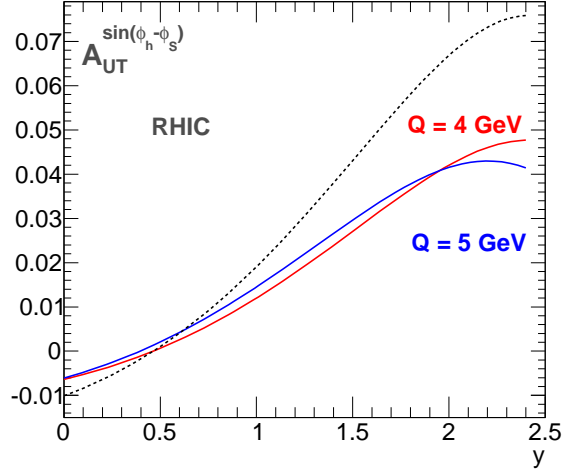}
\caption{Predictions for the Sivers single spin asymmetries
of Drell-Yan lepton pair production at RHIC, $\sqrt{S}=200$GeV,
as functions of rapidity for two different mass ranges. As a comparison,
we also show the prediction without the evolution effects for $Q=4$GeV case
as dotted line.}
\label{fig:rhic}
\end{figure}

Having constrained quark Sivers functions from HERMES/COMPASS
experiments, we will be able to make predictions for the SSAs in the Drell-Yan
processes with the evolution effects.
In Fig.~3, we show that for RHIC experiment at $\sqrt{S}=200$GeV,
as function of rapidity with $P_\perp$ integrated up to $2$GeV. We have flipped the sign for the quark Sivers function
because of the nontrivial universality property for the Sivers function.
For comparison, we have also plotted the prediction without the evolution
effects by setting ${\cal S}_{Sud}=0$ in Eq.~(\ref{sud}). From this, we see that the evolution
reduces the asymmetry by about a factor of 2. This is different from that
in Ref~\cite{Aybat:2011ta}, where an order of magnitude reduction
was indicated for the typical Drell-Yan experiments.

\begin{figure}[tbp]
\centering
\includegraphics[width=6cm]{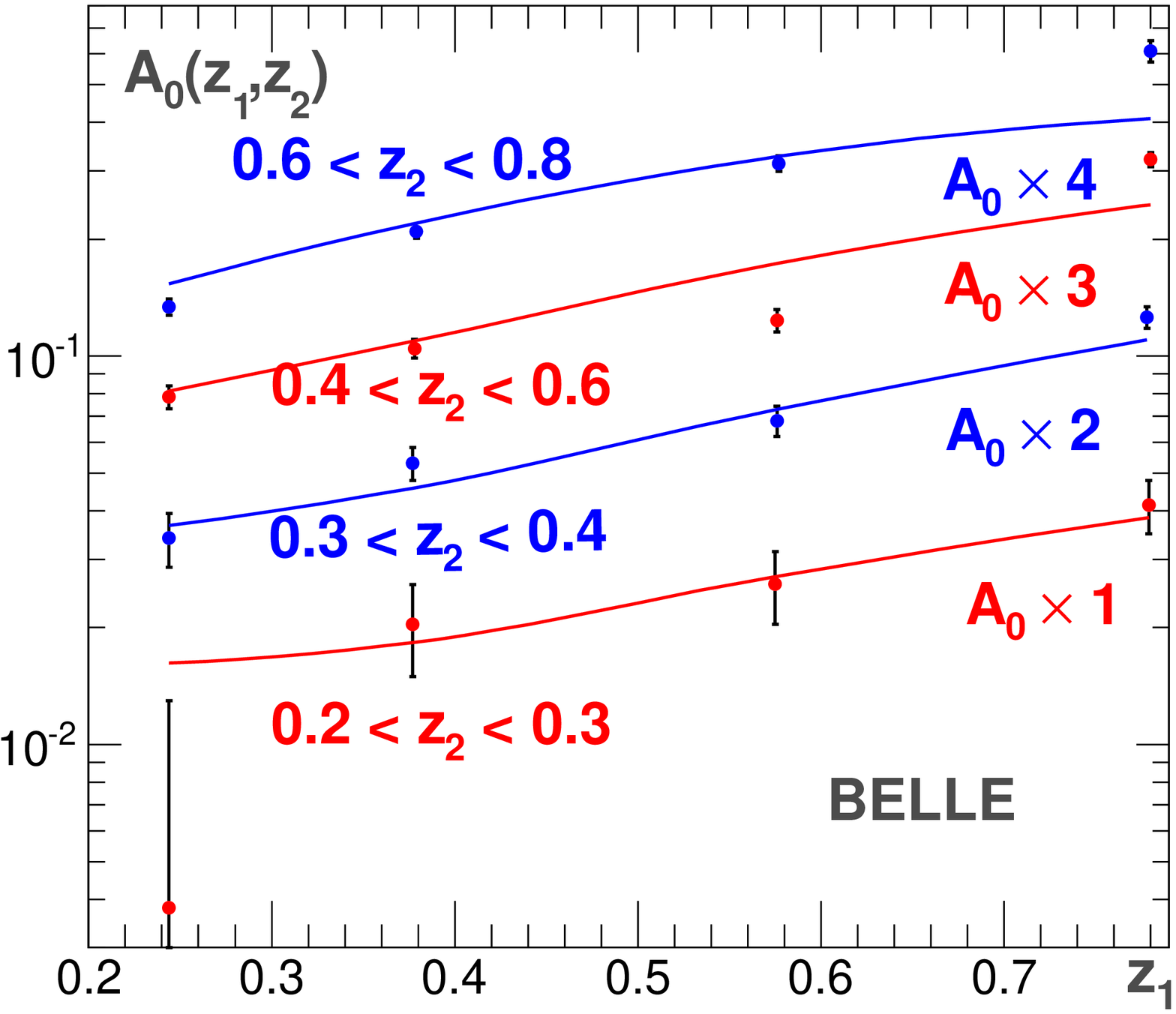}
\includegraphics[width=6cm]{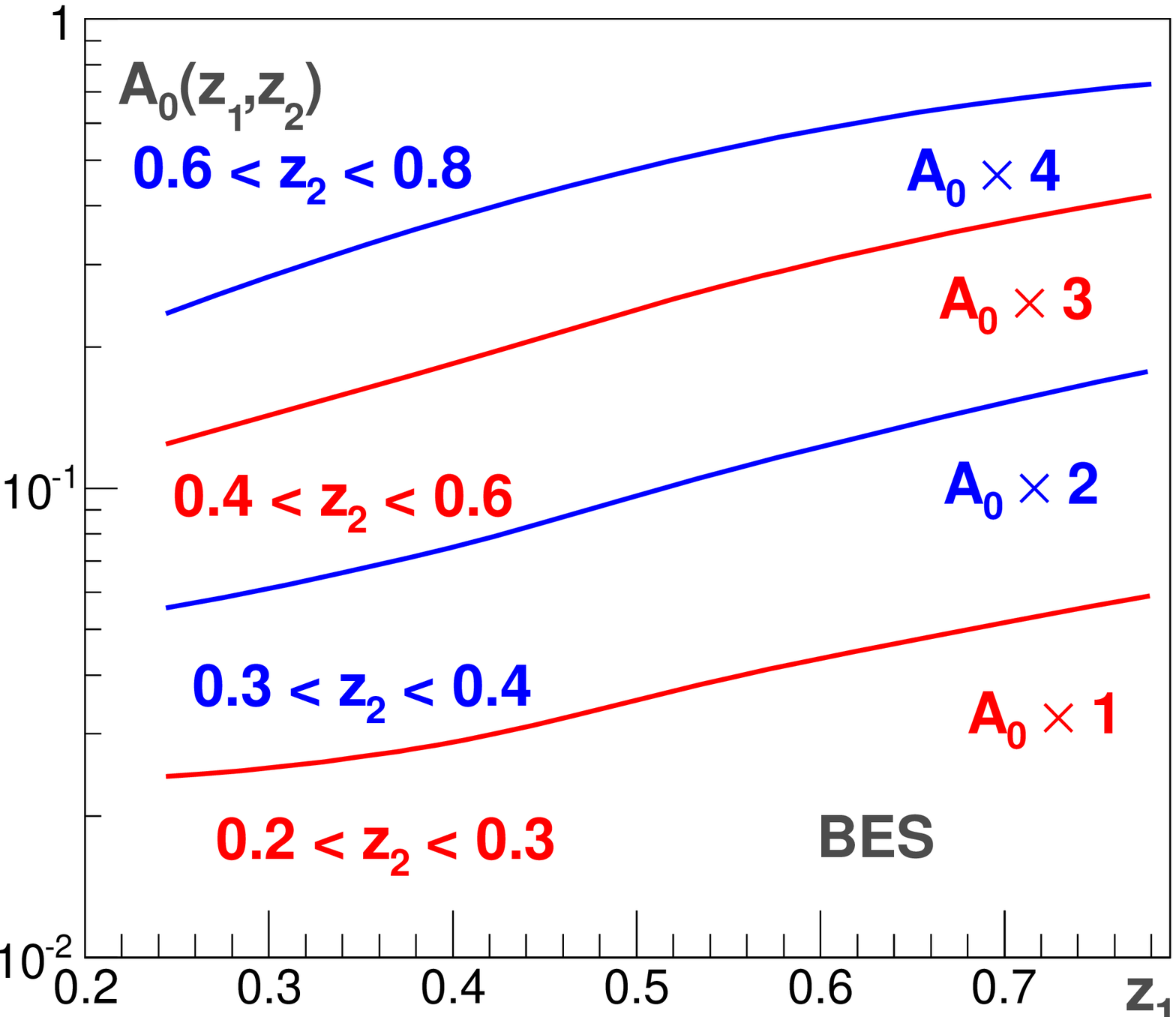}
\caption{The Collins asymmetries in di-hadron azimuthal angular
distributions in $e^+e^-$ annihilation processes: fit to the BELLE
experiment at $\sqrt{S}=10.6$GeV Ref.~\cite{Seidl:2008xc}, and predictions for the experiment
at BEPC at $\sqrt{S}=4.6$GeV.}
\label{fig:collins}
\end{figure}

We have done a number of cross checks for the above evolution effects.
First, we can tune the parameter in the calculations of Ref.~\cite{Aybat:2011ta}
to reproduce the $P_\perp$ spectrum of the Drell-Yan data, which leads to a much smaller $g_2=0.09$.
With that change, we can describe both SIDIS and Drell-Yan data in Fig.~1,
and the predicted SSA would be in the similar range as ours in Fig.~3.
Second, we determine the transverse momentum moment of
the quark Sivers function (Qiu-Sterman matrix element) from the fit in
Fig.~2, and calculate the SSA in Drell-Yan process by using the resummation
formula in Ref.~\cite{Kang:2011mr}, neglecting the scale dependence of the integrate
parton distributions and correlation functions and assuming the non-perturbative form
factor from Ref.~\cite{Konychev:2005iy} for Drell-Yan process with a mild $x$-dependence.
Again, we obtain the prediction in a similar range as that in Fig.~3.
In particular, this method provides an important step to matching the SIDIS
to Drell-Yan and W/Z boson productions in $pp$ collisions.

Finally, we turn to the energy evolution of the Collins asymmetries. 
We perform an analysis of the Collins asymmetries in the di-hadron azimuthal angular 
asymmetries in $e^+e^-$ annihilation from BELLE experiment~\cite{Seidl:2008xc}. Again, we parameterize
the Collins function at low energy scale $Q_0^2=2.4GeV^2$ as
$\widetilde H_{1}^{\perp\alpha}(z,b_\perp)=\left(\frac{-ib_\perp^\alpha M}{2z}\right)e^{ -{(g_h-g_c)  b^2}/z^2}N_q z^{\alpha_q} (1-z)^{\beta_q}
 \frac{(\alpha_q+\beta_q)^{\alpha_q+\beta_q}}{\alpha_q^{\alpha_q} \; \beta_q^{\beta_q}}D_q(z)$
and take the evolution effects of Eqs.~(\ref{sud}). The overall fit is very good,
as we show the comparison between the theory predictions and the BELLE
data. The fitting parameters for the Collins functions are listed in Table II,
with a $\chi^2/d.o.f=1.22$.
The combined analysis of the Collins asymmetries in $e^+e^-$ annihilation
and SIDIS leads to a consistent result.

%
\begin{table}[th!]
\caption{\label{tab:para} The fitting parameters $\{a_i^0\}$
for the Collins function at the input scale $Q^2=2.4\,\mathrm{GeV}$.}
\begin{ruledtabular}
\begin{tabular}{ccccc}
flavor $i$ &$N_i$ & $\alpha_i$ & $\beta_i$ & $g_c$ (GeV$^2$) \\
\hline
$u$   &  0.34$\pm$0.006   & 3.9$\pm$0.71    & 0.85$\pm$0.29     & 0.013$\pm$0.002     \\
$d$   & -0.34$\pm$0.013   & 0.4$\pm$0.31    & 0.31$\pm$0.41      & 0.013$\pm$0.002     \\
\end{tabular}
\end{ruledtabular}
\end{table}

An important feature we found in this fit is that both favored and disfavored
Collins functions saturates the positivity bounds. Therefore, it is very important to
check the energy dependence in other experiments. One idea place is the planed
electron-ion colliders~\cite{Boer:2011fh}, where SIDIS processes with wide
$Q^2$ coverage are the major focuses in the
proposal. Another place is the $e^+e^-$ annihilation process at different
energies. We suggest to investigate the di-hadron azimuthal correlation in
$e^+e^-$ annihilation at the BEPC of IHEP, Beijing, which can reach to
the center of mass energy around $\sqrt{S}=4.6$GeV. We show the prediction
for that energy in right panel of Fig.~3. Earlier experiments at SLAC around the
similar energy range demonstrated applicability of perturbative QCD description
of the jet structure~\cite{Hanson:1975fe}, which shall support to pursue similar studies at BEPC including the
Collins asymmetries. The initial state radiation events at BELLE can also
be used to study this asymmetry in various energies~\cite{Vossen}.

{\it Summary.}
In this paper, we have investigated the energy scale dependence
of the spin and azimuthal angular asymmetries in hard scattering processes.
We applied the Collins-Soper-Sterman
evolution at one-loop order in the moderate energy range,
which can well describe the transverse momentum spectrum
in existing SIDIS and Drell-Yan data.
We focused on the energy dependence of the Sivers and
Collins asymmetries in these processes, and performed a combined analysis
with the existing experimental data, including HERMES,
COMPASS, and BELLE experiments. The non-perturabtive TMD Sivers
function and Collins fragmentation functions are determined. The
predictions for future experiments are also presented. These experiments
will provide an important test for the TMD universality and strong interaction dynamics.
We will present more detailed results of this calculation, and the matching
from SIDIS to Drell-Yan and W/Z boson production in $pp$
collisions in a separate publication.

We thank J.~Collins for many stimulating discussions, suggestions, and
critical comments during the process of this project.
We thank M.~Anselmino, D.~Boer, J.~Qiu, A.~Prokudin, T.~Rogers, W.~Vogelsang,
and A.~Vossen for comments and correspondences.
We also thank Paul Hoyer for bringing us attention of Ref.~\cite{Hanson:1975fe}.
This work was partially supported by the U. S. Department of Energy via grant
DE-AC02-05CH11231.



\begin{thebibliography}{99}

\bibitem{Boer:2011fh}
  D.~Boer 
  {\it et al.},
  arXiv:1108.1713 [nucl-th];
  A.~Accardi
  {\it et al.},
  arXiv:1212.1701 [nucl-ex].

\bibitem{Brodsky:2002cx}
S.~J.~Brodsky, D.~S.~Hwang and I.~Schmidt,
Phys.\ Lett.\ B {\bf 530}, 99 (2002);
Nucl.\ Phys.\ B {\bf 642}, 344 (2002).


\bibitem{Collins:2002kn}
J.~C.~Collins,
Phys.\ Lett.\ B {\bf 536}, 43 (2002).




\bibitem{Airapetian:2009ae}
  A.~Airapetian {\it et al.}  [HERMES Collaboration],
  Phys.\ Rev.\ Lett.\  {\bf 103}, 152002 (2009).

\bibitem{Airapetian:2010ds}
  A.~Airapetian {\it et al.}  [HERMES Collaboration],
  Phys.\ Lett.\ B {\bf 693}, 11 (2010).

\bibitem{Alekseev:2008aa}
  M.~Alekseev {\it et al.}  [COMPASS Collaboration],
  Phys.\ Lett.\ B {\bf 673}, 127 (2009).

\bibitem{Adolph:2012sn}
  C.~Adolph {\it et al.}  [COMPASS Collaboration],
  Phys.\ Lett.\ B {\bf 717}, 376 (2012)
  [arXiv:1205.5121 [hep-ex]].

\bibitem{Adolph:2012sp}
  C.~Adolph {\it et al.}  [COMPASS Collaboration],
  Phys.\ Lett.\ B {\bf 717}, 383 (2012).



\bibitem{Seidl:2008xc}
  R.~Seidl {\it et al.}  [Belle Collaboration],
  Phys.\ Rev.\ D {\bf 78}, 032011 (2008)
  [Erratum-ibid.\ D {\bf 86}, 039905 (2012)].


\bibitem{Collins:1981uk}
J.~C.~Collins and D.~E.~Soper,
Nucl.\ Phys.\ B {\bf 193}, 381 (1981) [Erratum-ibid.\ B {\bf 213},
545 (1983)];
Nucl.\ Phys.\ B {\bf 197}, 446 (1982).


\bibitem{Collins:1984kg}
J.~C.~Collins, D.~E.~Soper and G.~Sterman,
Nucl.\ Phys.\ B {\bf 250}, 199 (1985).

\bibitem{Boer:2001he}
  D.~Boer,
  Nucl.\ Phys.\ B {\bf 603}, 195 (2001);
  Nucl.\ Phys.\ B {\bf 806}, 23 (2009);
  arXiv:1304.5387 [hep-ph].

\bibitem{Ji:2004wu}
  X.~Ji, J.~P.~Ma and F.~Yuan,
  Phys.\ Rev.\ D {\bf 71}, 034005 (2005);
Phys.\ Lett.\ B {\bf 597}, 299 (2004).

\bibitem{Idilbi:2004vb}
  A.~Idilbi, X.~Ji, J.~P.~Ma and F.~Yuan,
  Phys.\ Rev.\  D {\bf 70}, 074021 (2004).


\bibitem{Collins}
J.C.Collins, {\it Foundations of Perturbative QCD}, Cambridge University Press, Cambridge, 2011.

\bibitem{Aybat:2011zv}
  S.~M.~Aybat and T.~C.~Rogers,
  Phys.\ Rev.\ D {\bf 83}, 114042 (2011).

\bibitem{Aybat:2011ge}
  S.~M.~Aybat, J.~C.~Collins, J.~-W.~Qiu and T.~C.~Rogers,
  Phys.\ Rev.\ D {\bf 85}, 034043 (2012).

\bibitem{Landry:2002ix}
  F.~Landry, R.~Brock, P.~M.~Nadolsky and C.~P.~Yuan,
  Phys.\ Rev.\ D {\bf 67}, 073016 (2003);
  Phys.\ Rev.\ D {\bf 63}, 013004 (2001).

\bibitem{Aybat:2011ta}
  S.~M.~Aybat, A.~Prokudin and T.~C.~Rogers,
  Phys.\ Rev.\ Lett.\  {\bf 108}, 242003 (2012).

\bibitem{Anselmino:2012aa}
  M.~Anselmino, M.~Boglione and S.~Melis,
  Phys.\ Rev.\ D {\bf 86}, 014028 (2012).

\bibitem{Kang:2011mr}
  Z.~-B.~Kang, B.~-W.~Xiao and F.~Yuan,
  Phys.\ Rev.\ Lett.\  {\bf 107}, 152002 (2011).



\bibitem{Anselmino:2005ea}
  M.~Anselmino, {\it et al.},
  Phys.\ Rev.\ D {\bf 72}, 094007 (2005)
  [Erratum-ibid.\ D {\bf 72}, 099903 (2005)].

\bibitem{Schweitzer:2010tt}
  P.~Schweitzer, T.~Teckentrup and A.~Metz,
  Phys.\ Rev.\ D {\bf 81}, 094019 (2010).

\bibitem{deFlorian:2007aj}
  D.~de Florian, R.~Sassot and M.~Stratmann,
  Phys.\ Rev.\ D {\bf 75}, 114010 (2007).


\bibitem{Airapetian:2012ki}
  A.~Airapetian {\it et al.}  [HERMES Collaboration],
  arXiv:1212.5407 [hep-ex].


\bibitem{Ito:1980ev}
  A.~S.~Ito, {\it et al.},
  Phys.\ Rev.\ D {\bf 23}, 604 (1981).

\bibitem{Konychev:2005iy}
  A.~V.~Konychev and P.~M.~Nadolsky,
  Phys.\ Lett.\ B {\bf 633}, 710 (2006).
\bibitem{Hanson:1975fe}
  G.~Hanson, {\it et al.},
  Phys.\ Rev.\ Lett.\  {\bf 35}, 1609 (1975).

  \bibitem{Vossen} A.~Vossen, private communications.


\end{thebibliography}
\end{document}